\documentstyle[preprint,aps,psfig]{revtex}
\begin{document}
\hsize\textwidth\columnwidth\hsize\csname@twocolumnfalse\endcsname

\draft

\title{Spin Relaxation of Conduction Electrons}

\author{J. Fabian and S. Das Sarma}

\address{Department of
Physics, University of Maryland, College Park, MD
20742-4111}

\maketitle

\begin{abstract}
Prospect of building electronic devices in which electron spins
store and transport information has revived interest in the
spin relaxation of conduction electrons. Since spin-polarized
currents cannot flow indefinitely, basic spin-electronic devices
must be smaller than the distance electrons diffuse without losing
its spin memory. Some recent experimental and theoretical effort has
been devoted to the issue of modulating the spin relaxation. It has
been shown, for example, that in certain materials doping, alloying,
or changing dimensionality can reduce or enhance the spin relaxation
by several orders of magnitude. This brief review presents these
efforts in the perspective of the current understanding of the spin
relaxation of conduction electrons in nonmagnetic semiconductors 
and metals.
\end{abstract}
\vspace{2em}
\tighten
\newpage

\section{INTRODUCTION}

Electron spin is becoming increasingly popular in electronics.
New devices, now generally referred to as {\it spintronics}, exploit
the ability of conduction electrons in metals and
semiconductors to carry spin polarized current. Three factors
make spin of conduction electrons attractive for future technology:
(1) electron spin can store information, (2) the spin
(information) can be transferred as it is attached to
mobile carriers, and (3) the spin (information) can be
detected. In addition, the possibility of having long spin relaxation
time or spin diffusion length in electronic materials makes 
{\it spintronics} a viable potential technology.

Information can be stored in a system of electron spins because
these can be polarized. To represent bits, for example, spin up may
stand for one, spin down for zero. But the sheer existence of two
spin polarizations is of limited use if we do not have means of
manipulating them.  Currently used methods of polarizing electron
spins include magnetic field, optical orientation, and spin
injection.  Polarization by magnetic field is the traditional method
that works for both metals and semiconductors. Spin dynamics in
semiconductors, however, is best studied by optical orientation where
spin-polarized electrons and holes are created by a circularly
polarized light. Finally, in the spin injection technique a
spin-polarized current is driven, typically from a ferromagnet, into
the metallic sample.  Since spin is both introduced and transferred by
current, this method is most promising for spintronics.
Unfortunately, thus far spin injection has not been convincingly
demonstrated in semiconductors.

The second factor, the ability of information transfer by electron spins, 
relies
on two facts. First, electrons are mobile and second,
electrons have a relatively large spin memory. Indeed, conduction
electrons ``remember" their spins much longer
than they remember momentum
states.  In a typical metal, momentum coherence is lost after ten 
femtoseconds,
while spin coherence can survive more than a nanosecond. As a result, the
length $L_1$, the spin diffusion length, over which electrons remain 
spin polarized is much
longer than the mean free path distance $\ell$ over which their momentum 
is lost. Since
$L_1$ is the upper limit for the size of spintronic elements (in
larger elements the spin-encoded information fades away) it is not
surprising that significant effort went into finding ways of
reducing the spin relaxation.  Quite unexpectedly, in quantum wells,
but even in bulk semiconductors, donor doping was found to increase
the spin memory of conduction electrons by up to three orders of 
magnitude. In
metals one has much less freedom in manipulating electron states. A
theoretical study, however, predicts that even there spin memory can
be changed by orders of magnitude by band-structure tailoring.
Alloying of polyvalent metals with monovalent ones can increase
the spin memory by a decade or two. The ability of conduction
electrons to transport spin-polarized current over distances
exceeding micrometers has now been demonstrated in both metals and
semiconductors.

Finally, after the spin is transferred, it has to be
detected.  In many experiments the spin polarization is read
optically: photoexcited spin-polarized electrons
and holes in a semiconductor recombine by emitting circularly
polarized light; or the electron spins interact with
light and cause a rotation of the light polarization plane. 
It was discovered, however, that spin can be also
measured electronically, through charge-spin coupling. When spin
accumulates on the conductor side at the interface of a conductor and
a ferromagnet, a voltage or a current appears. By measuring the
polarity of the voltage or the current, one can tell the spin
orientation in the conductor.  Like spin injection, spin-charge
coupling has been demonstrated only in metals.

The operational synthesis of spin (information) storage, transfer,
and detection can be illustrated on concrete devices.  Spin
transistor is a trilayer that consists of a nonmagnetic metal (base)
sandwiched between two ferromagnets (emitter and collector).
Spin-polarized current injected into the base from the emitter causes
spin accumulation at the base-collector interface.  If the collector
magnetic moment is opposite to the spin polarization of the current
(and parallel to the emitter magnetic moment, if the injected
electrons are from the spin-minority subband), the
current flows from the base into the collector.  If the collector
magnetic moment is parallel to the spin polarization, the current is
reversed.  In order for the spin accumulation to occur, the current
in the metallic base must remain polarized--the base must be thinner
than $L_1$.  Similar principles work in the giant magnetoresistance
effect.  Multilayer structures with alternating nonmagnetic and
ferromagnetic metals have their resistance strongly dependent on the
relative orientation of the ferromagnetic moments.  The resistance is
small if the moments point in the same directions, and large if the
directions of neighboring moments are reversed.  Again, the
information about the moment of a ferromagnetic layer is encoded into
electron spins which carry this information through a contiguous
nonmagnetic metal into another ferromagnet. Here the information is
read and in ideal case the electron is let into the ferromagnet  only
if its spin is opposite to the direction of the ferromagnetic moment.
Otherwise the electron is scattered at the interface.

Several recent reviews focus on spin-polarized transport.
An overview of the subject can be found in \cite{prinz95}.
Spin transistors, spin injection, and charge-spin coupling
in metallic systems
is treated in \cite{johnson93a}; a comprehensive account
of optical orientation is given in \cite{meier84}, and
recent reviews of giant magnetoresistance are in 
\cite{ansermet98,allen97}.
Many suggested spintronic devices have not been
demonstrated yet, but their potential seems enormous.
Industrial issues related to spintronics can be found in
\cite{bond93,simonds95}, and \cite{gregg97} describes some of
the recent spintronic schemes and devices.

The present article introduces basic concepts of the spin relaxation
of conduction electrons and identifies important unresolved issues in
both semiconductors and metals. Particular emphasis is given to the
recent experimental and theoretical work that attempts to enhance
and/or understand electron spin coherence in electronic materials.

\section{MECHANISMS OF SPIN RELAXATION}

Spin relaxation refers to the processes that bring an
unbalanced population of spin states into equilibrium.
If, say, spin up electrons are injected into a metal
at time $t=0$ creating a spin imbalance,
at a later time, $t=T_1$ (the so called spin relaxation time),
the balance is restored by a coupling between spin and
orbital degrees of freedom. Three spin-relaxation mechanisms
have been found to be relevant for conduction electrons (Fig. \ref{fig:1}):
the Elliott-Yafet, D'yakonov-Perel', and Bir-Aronov-Pikus.

The Elliott-Yafet mechanism is based on the fact
that in real crystals Bloch states are not spin eigenstates. Indeed,
the lattice ions induce the spin-orbit interaction that mixes the
spin up and spin down amplitudes\cite{elliott54}. Usually the
spin-orbit coupling $\lambda$ is much smaller than a typical band
width $\Delta E$ and can be treated as a perturbation. Switching the
spin-orbit interaction adiabatically, an initially spin up (down)
state acquires a spin down (up) component with amplitude $b$ of order
$\lambda/\Delta E$.  Since $b$ is small, the resulting states can
be still named ``up'' and ``down'' according to their
largest spin component.  Elliott  noticed \cite{elliott54} that an
ordinary (spin independent) interaction with impurities, boundaries, 
interfaces, and phonons can connect ``up'' with
``down'' electrons, leading to spin relaxation whose rate $1/T_1$
is proportional to $b^2/\tau$ ($\tau$ being
the momentum relaxation time determined by ``up'' to ``up''
scattering).  Additional spin-flip scattering is provided by
the spin-orbit interaction of impurities, and by the phonon-modulated
spin-orbit interaction of the lattice ions 
(Overhauser \cite{overhauser53}).
The latter should be taken together with the Elliott phonon
scattering to get the correct low-temperature behavior of $1/T_1$
\cite{yafet63}.  Yafet showed \cite{yafet63} that $1/T_1$ follows the
temperature dependence of resistivity: $1/T_1\sim T$ at temperatures
$T$ above the Debye temperature $T_D$, and $1/T_1\sim T^5$ at very
low $T$ in clean samples (neutral impurities lead to $T$-independent
spin relaxation). Elliott-Yafet processes due to the electron-electron 
scattering in semiconductors were evaluated by Boguslawski  
\cite{boguslawski80}.

In crystals that lack inversion symmetry (such as zincblende
semiconductors) the spin-orbit interaction lifts the spin degeneracy:
spin up and spin down electrons have different energies even when
in the same momentum state. This is equivalent to having
a momentum-dependent internal magnetic field  ${\bf B}({\bf k})$
which is capable of flipping spins through the interaction term like
${\bf B}({\bf k})\cdot {\bf S}$, with ${\bf S}$ denoting the electron
spin operator. (This term can be further modulated by strain or by
interface electric fields).  D'yakonov and Perel'
showed that the lifting of the spin degeneracy leads to spin
relaxation\cite{dyakonov71}.  Typically the distance between spin up
and down bands is much smaller than the frequency $1/\tau$ of
ordinary scattering by impurities, boundaries, or phonons.  
Consider an electron
with momentum ${\bf k}$. Its spin precesses along the axis given by
${\bf B}({\bf k})$.  Without going the full cycle, the electron
scatters into momentum ${\bf k}'$ and begins to precess along the
direction now given by ${\bf B}({\bf k}')$, and so on. The electron
spin perceives the scattering through randomly changing precession
direction and frequency. The precession angle along the axis
of initial polarization (or any other fixed axis) diffuses so its
square becomes about $(t/\tau)(\omega\tau)^2$ after time $t$
($\omega$ is the typical precession frequency). By definition $T_1$
is the time when the precession angle becomes of order one. Then
$1/T_1 \approx \omega (\omega\tau)$. The factor $(\omega\tau)$ is a
result of motional narrowing as in nuclear magnetic resonance
\cite{kittel}.  The spin relaxation {\it rate} $1/T_1$ is
proportional to the momentum relaxation {\it time} $\tau$.  We note
that in strong magnetic fields the precession along the randomly
changing axis is suppressed (spins precess along the external
field\cite{dyakonov73} and electron cyclotron motion averages over
different internal magnetic fields\cite{ivchenko73,zakharchenya76}),
leading to a reduction of the D'yakonov-Perel' spin relaxation.

Another source of spin relaxation for conduction electrons was
found by Bir, Aronov, and Pikus \cite{bir76} in the electron-hole
exchange interaction. This interaction depends on the spins of
interacting electrons and holes and acts on electron spins as some
effective magnetic field. The spin relaxation takes place as electron
spins precess along this field. In many cases, however, hole spins
change with the rate that is much faster than the precession
frequency. When that happens the effective field which is generated
by the hole spins fluctuates and the precession angle about a
fixed axis diffuses as in the case of the D'yakonov-Perel' process.
The electron spin relaxation rate $1/T_1$ is then
``motionally'' reduced and is proportional to the hole spin
relaxation time.  Similar reduction of $1/T_1$ occurs if holes that
move faster than electrons change their momentum before electron
spins precess a full cycle\cite{bir76}.  The Bir-Aronov-Pikus spin
relaxation, being based on the electron-hole interaction, is relevant
only in semiconductors with a significant overlap between electron
and hole wave functions.

\section{SEMICONDUCTORS}

Spin relaxation in semiconductors is rather complex.
First, there are different charge carriers to consider. Both
electrons and holes can be spin polarized and carry spin-polarized
currents.  Furthermore, some features of the observed luminescence
polarization spectra must take into account excitons, which too, can
be polarized.  Second, in addition to temperature and impurity
content the spin relaxation is extremely sensitive to factors like
doping, dimensionality, strain, magnetic and electrical fields.  The
type of dopant is also important:  electrons in $p$-type samples, for
example, can relax much faster than in $n$-type samples. And,
finally, since the relevant electron and hole states are typically
very close to special symmetry points of the Brillouin zone,
subtleties of the band structure often play a decisive role in
determining which spin relaxation mechanism prevails. (Band
structure also determines what is polarized--often due to degeneracy
lifting,  spin and orbital degrees are entirely mixed and the total
angular momentum is what is referred to as ``spin.'') The above
factors make sorting out different spin relaxation mechanisms a
difficult task.

The first measurement of $T_1$ of free carriers in a semiconductor
was reported in Si by Portis et al. \cite{portis53} and Willenbrock
and Bloembergen \cite{willenbrock53}; these measurements were done by
conduction electron spin resonance. Silicon, however,
remains still very poorly understood in regards to its spin transport
properties.  Very little is known, for example, about electronic
spin-flip scattering by conventional $n$ and $p$ dopants. Considering
that Si may be an important element for spintronics since it is
widely used in conventional electronics, its spin relaxation
properties should be further investigated.

Much effort was spent on  III-V semiconductors where optical
orientation \cite{meier84} enables direct measurement of $T_1$.
In these systems holes relax much faster than electrons because
hole Bloch states are almost an equal admixture of  spin up and
down eigenstates. The Elliott-Yafet mechanism
then gives $T_1$ of the same order as $\tau$. In quantum wells (QW),
however, $T_1$ of holes was predicted by Uenoyama and
Sham\cite{uenoyama90} and Ferreira and Bastard \cite{ferreira91} to
be quenched, and even longer than the electron-hole recombination time. 
This was observed experimentally in $n$-modulation
doped GaAs QWs by Damen et al.  \cite{damen91} who measured  hole
spin relaxation time of 4 ps at 10 K. Hole and exciton spin
relaxation was reviewed by Sham\cite{sham93}.

Compared to holes, electrons in III-V systems remember their spins
much longer and are therefore more important for spintronic
applications.  Typical measured values of electron $T_1$ range from
$10^{-11}$ to $10^{-7}$s. All the three spin relaxation mechanisms
have been found contributing to $T_1$. Although it is difficult to
decide which mechanism operates under the specific experimental
conditions (this is because in some cases two mechanisms yield
similar $T_1$, but also because experiments often disagree with each
other\cite{zerrouati88}), some general trends are followed. The
Elliott-Yafet mechanism dominates in narrow-gap semiconductors,
where $b^2$ is quite large ($\Delta E\approx E_g$ is small).
Chazalviel\cite{chazalviel75} studied $n$-doped InSb ($E_g\approx 0.2
$ eV) and found that Elliott-Yafet scattering by ionized impurities
explains the observed $1/T_1$.

If band gap is not too small, the D'yakonov-Perel' mechanism has been
found relevant at high temperatures and sufficiently low densities of
holes. The D'yakonov-Perel' mechanism can be quite easily
distinguished from the Elliott-Yafet one: the former leads
to $1/T_1\sim \tau$ while for the latter $1/T_1 \sim 1/\tau$. The
increase in the impurity concentration decreases the efficiency of
the D'yakonov-Perel' processes and increases those due to Elliott and
Yafet.  Another useful test of the D'yakonov-Perel' mechanism is its
suppression by magnetic field\cite{ivchenko73,zakharchenya76}. The
first experimental observation of the D'yakonov-Perel' mechanism was
reported by Clark et al.  on moderately doped $p$ samples of
GaAs\cite{clark76} and GaAlAs\cite{clark75}. Later measurements on
less doped samples of GaAs by Maruschak et al.\cite{maruschak83} and
Zerrouati et al.\cite{zerrouati88} confirmed that the
D'yakonov-Perel' mechanism is dominant in GaAs at elevated
temperatures.

At low temperatures and in highly $p$-doped samples (acceptor
concentration larger than $10^{17}$ cm$^{-3}$) the Bir-Aronov-Pikus
mechanism prevails. As the acceptor concentration increases this
mechanism reveals itself at progressively higher temperatures.
An increase of $1/T_1$ with increasing $p$ doping signals
that the electron-hole spin relaxation is relevant.
This was demonstrated in $p$-type GaAs (for example, Zerrouati et al.  
\cite{zerrouati88}, Maruschak et al. \cite{maruschak83},
and Fishman and Lampel\cite{fishman77}) and GaSb (Aronov et al.
\cite{aronov83}).  The physics of spin relaxation
in $p$-doped III-V semiconductors is very rich because several 
different mechanisms have been shown relevant. More work, 
however, still needs to be done. It is not clear, for example, 
what happens at very low temperatures and in very pure 
samples\cite{zerrouati88}.  There are some indications that 
at very low temperatures both the D'yakonov-Perel'  
and the Bir-Aronov-Pikus mechanisms can explain the observed data 
at whatever doping\cite{zerrouati88}. Excellent reviews
 of conduction electron spin relaxation in bulk III-V 
semiconductors are \cite{pikus84,pikus88}. These references 
contain both experimental data and many useful formulas of $1/T_1$.

Electron spin relaxation has been studied also in quantum wells.
That spin dynamics in quantum wells differs from that in the bulk is
obvious from the fact that the relevant spin relaxation mechanisms
are very sensitive to factors like mobility (which is higher in QWs),
electron-hole separation (smaller in QWs) and electronic band
structure (more complicated in QWs  because of subband structures
and interface effects). Furthermore,  the quality of QW samples
is very important since $1/T_1$ is strongly influenced by
localization and defects. The first measurement of conduction
electron $T_1$ in a QW was reported by Damen et al. \cite{damen91}
who studied the dynamics of luminescence polarization in
$p$-modulation doped GaAs/AlGaAs, and obtained $T_1\approx 0.15 $ ns
at low temperatures. This relaxation time is three to four
times smaller than in a similar bulk sample (the acceptor concentration was
$4\times10^{11}$ cm$^{-2}$).  It was concluded\cite{damen91} that the
relevant mechanism was Bir-Aronov-Pikus. The recent theoretical study
by Maialle and Degani \cite{maialle97} of the Bir-Aronov-Pikus
relaxation in QWs indicates that, to the contrary, this
mechanism is not efficient enough to explain the experiment. Another
possibility is the D'yakonov-Perel' mechanism.  Bastard and
Ferreira\cite{bastard92} calculated the effectiveness of
this mechanism for the case of ionized impurity scattering.  Their
calculation shows\cite{sham93} that the D'yakonov-Perel'
mechanism also too weak to explain the experiment. Although some
assumptions of the theoretical studies may need to be reexamined
(the major difficulty seems to be estimating $\tau$) \cite{sham93}, further
experimental work (such as temperature and doping dependence) is
required to decide on the relevant mechanism. Recently Britton
et al. \cite{britton98} studied the spin relaxation in undoped 
GaAs/AlGaAs multiple quantum wells at room temperature. The measured 
relaxation times vary between 0.07 and 0.01 ns, decreasing strongly 
with increasing confinement energy. These results seem to be consistent
with the D'yakonov-Perel' mechanism \cite{britton98}. 

Spin relaxation studies in quantum wells also promise better 
understanding of interface effects. In an inversion layer
an electric field arises from the electrostatic confinement. 
This field induces a spin-orbit interaction which contributes 
to the spin-splitting (the so called Rashba splitting) of electron 
bands in addition to the inversion-asymmetry splitting. This should 
enhance the efficiency of the D'yakonov-Perel' mechanism. Spin 
precession of conduction electrons in GaAs inversion layers was 
investigated by Dresselhaus et al. \cite{dresselhaus92} using 
antilocalization. The spin relaxation was found to be due to the 
D'yakonov-Perel' mechanism, but the spin splitting was identified
(by magnetic field dependence) to be primarily due to the inversion 
asymmetry. This is consistent with an earlier theoretical study 
of Lommer et al.  
\cite{lommer88} of spin 
splitting in heterostructures, which predicted that in GaAs/AlGaAs QWs the 
Rashba term in the Hamiltonian is weak.  In narrow-band semiconductors,
however, Lommer et al. predict that the Rashba term becomes relevant.
But this remains a not-yet-verified theoretical 
prediction. Another
interesting study of the interface effects was done recently by 
Guettler et al.\cite{guettler98} following the calculations of 
Vervoort et al. \cite{vervoort97}. Quantum well systems in which 
wells and barriers have different host atoms (so called 
``no-common-atom'' 
heterostructures) were shown to have conduction electron spin 
relaxation enhanced by orders of magnitude compared to 
common-atom heterostructures. In particular, spin relaxation times
in (InGa)As/InP QWs were found to be 20 (90) ps for electrons (holes),
while the structures with common host atoms (InGa)As/(AsIn)As 
have spin relaxation times much longer: 600 (600) ps. 
This huge difference between otherwise 
similar samples is attributed to the large electric fields arising 
from the asymmetry at the interface (interface dipolar fields) 
\cite{guettler98}.

Spin relaxation of conduction electrons can be controlled. This was 
first realized by Wagner et al. \cite{wagner93} who $\delta$-doped
GaAs/AlGaAs double heterostructures with Be (as acceptor).
The measured spin relaxation time was about 20 ns which is {\it 
two} orders of magnitude longer than in similar homogeneously 
$p$-doped GaAs. The understanding of this finding is the following.
The sample was heavily doped ($8\times 10^{12}$ cm$^{-2}$) so the
Bir-Aronov-Pikus mechanism was expected to dominate the relaxation.
Photogenerated electrons, however, were spatially separated from 
holes which stayed mostly at the center of the GaAs layer, close to 
the Be dopants.  There was, however, still enough overlap 
between electrons and the holes for efficient recombination so that 
the radiation polarization could be studied.  The decrease of the 
overlap between electrons and holes reduced the efficiency of the 
Bir-Aronov-Pikus mechanism and increased $T_1$.  
This experiment can be also taken as a confirmation that the Aronov-Bir-Pikus 
mechanism is dominant in heavily $p$-doped heterostructures.

The next important step in controlling  spin relaxation was the
observation of  a large enhancement of the spin memory of electrons in
II-VI semiconductor QWs by Kikkawa et al. 
\cite{kikkawa97}.  Introducing a (two dimensional) electron gas by 
$n$-doping the II-VI QWs was found to increase electronic spin 
memory by several orders of magnitude. The studied samples were 
modulation-doped Zn$_{1-x}$Cd$_x$Se quantum wells with electron 
densities $2\times 10^{11}$ and $5\times 10^{11}$ cm$^{-2}$ (an
additional insulating sample was used as a benchmark).  
Spin polarization was 
induced by  a circularly polarized pump pulse directed normal to the 
sample surface.  The spins, initially polarized normal, began to 
precess along an external magnetic field oriented along the surface
plane. After a time $\delta t$, a probe pulse of a linearly polarized 
light detected the orientation of the spins. The major result of the 
study was that in doped samples electron spin remained polarized  for 
almost three orders of magnitude longer than in the insulating (no 
Fermi see) sample. The measured $T_1$ was on the nanosecond scale,
strongly dependent on the magnetic field and weakly dependent on 
temperature and mobility. Although the nanosecond time scales and the 
increase of the observed polarization in strong magnetic fields 
(usually a Tesla) could be explained by the D'yakonov-Perel' 
mechanism\cite{zerrouati88}, the temperature and mobility (in)dependence
remain a puzzle. The overall increase of $T_1$ by donor doping can
be understood in the following way\cite{kikkawa97}. In 
insulating samples photoexcited spin-polarized electrons quickly recombine 
with  holes. This happens in picoseconds. In the presence of a Fermi sea 
photoexcited electrons do not recombine (there are 
plenty other electrons available for 
recombination) so they lose their spins in nanoseconds, which 
are natural time scales for spin relaxation. There is a 
caveat, however. The above scenario is true only if holes lose their 
spins faster than they recombine with electrons. Otherwise only
electrons from the Fermi sea with a preferred spin would recombine, leaving 
behind a net opposite spin that counters that of the photoexcited 
electrons. The fast hole relaxation certainly happens in the bulk 
(and similar enhancement of $T_1$ has been observed in $n$-doped 
bulk GaAs by 
Kikkawa and Awschalom\cite{kikkawa98}), but not necessarily in 
quantum heterostructures\cite{uenoyama90,ferreira91,damen91}. This 
issue therefore remains open. Very recent optically pumped nuclear 
magnetic resonance measurements \cite{kuzma98} in $n$-doped 
AlGaAs/GaAs multiple quantum well systems indicate unusually
long $T_1 \agt 100 \mu$s at temperatures below 500 mK in the    
two-dimensional electron gas system under the application of a 
strong ($\agt 12$ T) external magnetic field. It is unclear whether
this remarkable decoupling (that is, $T_1 \agt 100 \mu$s) of the
two-dimensional electron gas spins from its environment is an
exotic feature of the fractional quantum Hall physics dominating
the system, or is a more generic effect which could be controlled
under less restrictive conditions.

It was recently demonstrated that spin polarized
current can flow in a semiconductor. H\"{a}gele et al. 
\cite{hagele98} used a simple but ingenuous setup that consisted of 
a micrometer $i$-GaAs block attached to a $p$-modulation doped GaInAs 
QW layer.  The free surface of the GaAs block was illuminated by a 
circularly polarized light. The photogenerated electrons then drifted 
towards the QW under the force of an applied electric field 
(photoexcited holes 
moved in the opposite direction towards the surface). The electrons 
recombined with holes upon hitting the QW, emitting light. By 
observing the polarization of the emitted light H\"{a}gele et al.
concluded that electrons captured by the QW were polarized. 
The spin was almost completely conserved after the electrons traveled 
as long as 4 micrometers and under the fields up to 6 kV/cm,
indicating very long spin diffusion lengths in these experiments 
\cite{hagele98}.

\section{METALS}

Only a dozen elemental metals have been investigated for spin
relaxation so far. Early measurements of
$T_1$ were done by the conduction electron spin resonance technique.
This technique was demonstrated for metals by Griswold et al.
\cite{griswold52}, and Feher and Kip \cite{feher55} used it to make
the first $T_1$ measurement of Na, Be, and Li. This and subsequent
measurements established that $1/T_1$ in metals depends strongly on
the impurity content (especially in light metals like Li and Be)
and grows
linearly with temperature at high temperatures. Typical spin
relaxation time scales were set to nanoseconds, although in very pure
samples $T_1$ can reach microseconds at low temperatures (for
example in sodium, as observed by Kolbe \cite{kolbe71}).
Reference \cite{monod79} is a good source of these early spin
relaxation measurements.

The next wave of measurements started with the realization of
spin injection in metals. Suggested theoretically by Aronov
\cite{aronov76b}, spin injection was first demonstrated in Al by
Johnson and Silsbee\cite{johnson85}. Later measurements were done on
Au\cite{johnson93c} and Nb\cite{johnson94b} films. The spin injection
technique enables measurements of $T_1$ in virtually no magnetic
fields so that $T_1$ can now be measured in
superconductors, spin glasses, or Kondo systems where magnetic field
qualitatively alters electronic states.  Furthermore, by eliminating
the need for magnetic fields to polarize electron spins one avoids
complications like inhomogeneous line broadening, arising from $g$
factor anisotropy.  Johnson also succeeded in injecting spin
polarized electrons into superconducting Nb films\cite{johnson94b}.
Spin relaxation of electrons (or, rather, quasiparticles) in
superconductors is, however, poorly understood and the experiments,
now done mostly on high-$\rm T_c$ materials \cite{vasko97,dong97}
only manifest the lack of theoretical comprehension of the subject. 
Still waiting for its demonstration is spin injection into 
semiconductors. Although
it was predicted long ago by Aronov and Pikus \cite{aronov76a} it
still remains a great experimental challenge.

The observations that $1/T_1\sim T$ at high temperatures, the strong
dependence on impurities,  and characteristic nanosecond time scales
has led to the general belief that  conduction electrons in metals
lose their spins by the Elliott-Yafet mechanism.  Although simple
estimates and even some analytical calculations
were done for simplest metals like Na (Yafet\cite{yafet63}),
careful numerical calculations are   lacking.  Experimental data are
usually analyzed to see if the simple relation suggested by
Yafet\cite{yafet63},
\begin{eqnarray} \label{eqn:yafet}
1/T_1 \sim b^2\rho,
\end{eqnarray}
where $\rho$ is  resistivity, is obeyed. The spin-mixing $b^2$ is the
fitting parameter so the temperature behavior of $1/T_1$ is
determined solely by $\rho$. At high temperatures $1/T_1\sim\rho\sim
T$ as observed.  At low temperatures the spin relaxation should obey
the Yafet law $1/T_1\sim T^5$ (in parallel to the Bloch law $\rho\sim
T^5$), but so far this has not been observed, mainly due to the large
contribution from impurity and boundary scattering.  (Even after
subtracting this temperature independent background the uncertainties
of the measurements prevent a definite experimental conclusion about
the low $T$ behavior.)

Equation \ref{eqn:yafet} suggests that dividing
$1/T_1$ by $b^2$ one obtains resistivity, up to a multiplicative
(material independent) constant. Resistivity, divided by its value
$\rho_D$ at $T_D$  and expressed as function of reduced temperature
$T/T_D$ follows a simple Gr\"{u}neisen curve, the same for all simple
metals.  Monod and
Beuneu\cite{monod79} applied this reasoning to then available
experimental data of $T_1$.  For the spin mixing $b^2$ they
substituted values obtained from atomic parameters of the
corresponding elements. The resulting (revised) scaling is reproduced
in Fig. \ref{fig:2}. (The original scaling \cite{monod79} has
$\Gamma_s$ divided by $b^2$, not by $b^2\rho_D$.) The picture is
surprising.  While some metals (the ``main group'') nicely follow a
single Gr\"{u}neisen curve, others do not. There seems to be no
obvious reason for the observed behavior.  Metals Na and Al, for
example, are quite similar in that their atomic $b^2$ differ by less
than 10\%\cite{monod79}.  Yet the spin relaxation times at $T_D$ are
0.1 ns for Al and 20 ns for Na \cite{fabian98}.

The solution to this puzzle can be found by
recognizing\cite{fabian98} that the main group is formed by {\it
monovalent} alkali and noble metals, while the metals with
underestimated $b^2$, Al, Pd, Mg, and Be are {\it polyvalent} (no
other metals have been measured for $T_1$ in a wide enough
temperature region). Monovalent metals have their Fermi surfaces
inside Brillouin zone boundaries so that distance between neighboring
bands, $\Delta E$ is quite uniform and of the order of the Fermi
energy $E_F$. The spin mixing is then $b^2\approx (\lambda/E_F)^2$
for all states on the Fermi surface.  Polyvalent metals, on the
other hand, have Fermi surfaces which cross Brillouin zone boundaries,
and often also special symmetry points and accidental degeneracy
lines. When this happens the electron spin relaxation is
significantly enhanced. This was first noted by Silsbee and Beuneu
\cite{silsbee83} who estimated the contribution to Al $1/T_1$
from accidental degeneracy lines.  Later the present authors
gave a rigorous and detailed treatment of how not only accidental
degeneracy, but all the band anomalies contribute to $1/T_1$
\cite{fabian98}. This treatment led to the spin-hot-spot model
\cite{fabian98} which explains why all the measured polyvalent metals
have spin relaxation faster than expected from a naive theory.
In addition to explaining experiment, the spin-hot-spot model
predicts the behavior of other polyvalent metals. The model is
illustrated in Fig. \ref{fig:3}.

As an example, consider a metal whose Fermi surface crosses a
single Brillouin zone boundary\cite{fabian98,fabian99}. The distance between
energy bands $\Delta E$ is about $E_F$ for all Fermi surface states
except those close to the boundary. There $\Delta E\approx
2V$ \cite{kittel}, where $V$ is the Fourier component of the  lattice
potential associated with the boundary. Since in most cases $V\ll E_F$
the spin mixing $b^2\approx (\lambda/V)^2$ is much larger than on
average. If an electron jumps into such states, the chance that
its spin will be flipped is much enhanced. Similarly if the electron
jumps {\it from} these ``spin hot spots.'' But how much the states with
$\Delta E \approx 2V$ contribute to spin relaxation depends on how
many they are relative to the number of states on the Fermi surface.
A single electron experiences thousands of jumps due to momentum
scattering before its spin flips. Therefore the spin relaxation rate
$1/T_1$ is determined by the average $\langle b^2\rangle$ of $b^2$
over the Fermi surface.  The majority of states with $\Delta E\approx
E_F$ contribute $(\lambda/E_F)^2\times 1$ (the value of $b^2$ times
the probability of occurrence, which in this case is close to one) to $\langle
b^2 \rangle$. The probability of finding a state with $\Delta E
\approx 2V$ on the Fermi surface turns out to be about $V/E_F$
\cite{fabian99}, so
the spin hot spots contribute about $(\lambda/V)^2\times (V/E_F)$, 
which is  $(\lambda/E_F)^2\times (E_F/V)$. This is larger by $E_F/V$
than the contribution from ordinary states. Typically $E_F/V\approx
10$, and considering that in reality the Fermi surface crosses
more than one Brillouin zone boundary, the spin relaxation can be
enhanced up to two orders of magnitude. Electron jumps that include
at least one spin-hot-spot state dominate spin relaxation to the
extent that the majority of scattering events (those outside the spin
hot spots) can be neglected.

The spin-hot-spot picture not only solves a long-standing
experimental puzzle, but also shows a way to tailor the spin
relaxation of electrons in a conduction band. Spin relaxation of
a monovalent metal, for example, can be enhanced by alloying with a
polyvalent metal. This brings more electrons into the
conduction band. As the Fermi surface increases, it begins to cross
Brillouin zone boundaries and other spin-hot-spot regions.
The enhancement of $1/T_1$ can be significant. Similarly, $1/T_1$
can be reduced by orders of magnitude by alloying polyvalent metals
with monovalent. Applying pressure, reducing the dimensionality,
or doping into a semiconductor conduction bands as well
as any other method of modifying the band structure should work. 
The rule of thumb
for reducing $1/T_1$ is washing the spin hot spots off the Fermi
surface. (Another possibility would be to inhibit scattering in or
out the spin hot spots, but this is hardly realizable.)

The most important work ahead is to catalog $1/T_1$ for more
metallic elements and alloys. So far only the simplest metals have
been carefully studied over large enough temperature ranges,
but even in these cases it is not clear, for example, as to how
phonon-induced $1/T_1$ behaves at low temperatures.  It is plausible
that understanding $1/T_1$ in the transition metals will require new
insights (such as establishing the role of the $s$-$d$ exchange). Another
exciting possibility is that the measurements at high enough
temperatures will settle the question of the so called ``resistivity
saturation''\cite{fisk76} which occurs in many transition metals. Indeed, the two
competing models of this phenomenon imply different
scenarios for $1/T_1$: the ``phonon ineffectiveness'' model\cite{cote78} implies
saturation of $1/T_1$, while the model emphasizing the role of 
quantum corrections
to Boltzmann theory\cite{chakraborty79} apparently does not\cite{allenp}.  
Finally, theory should
yield probabilities of various spin-flip processes
in different metals. Empirical pseudopotential and density functional
techniques seem quite adequate to perform such calculations. Some
work in this direction is already under way\cite{fabian}.

\section{conclusion}

We have provided a brief informal review of the current understanding
of spin relaxation phenomenon in metals and semiconductors. 
Although studying spin relaxation through electron spin resonance
measurements and developing its microscopic understanding through
quantitative band structure analyses were among the more active
early research areas in solid state physics (dating back to the early
1950s), it is surprising that our current understanding of the phenomenon
is quite incomplete and is restricted mostly to bulk elemental metals
and some of the III-V semiconductor materials (both bulk and quantum
well systems). There is a great deal of renewed current interest in 
the subject because of the potential spintronics applications offering
the highly desirable possibility of monolithic integration of
electronic, magnetic, and optical devices in single chips as well as
the exciting prospect of using spin as a quantum bit in proposed
quantum computer architectures. It should, however, be emphasized 
that all of these proposed applications necessarily require comprehensive
quantitative understanding of physical processes controlling spin
coherence in electronic materials. In particular, there is an acute
need to develop techniques which can manipulate spin dynamics in 
a controlled coherent way which necessitates having long spin
relaxation times and/or spin diffusion lengths. Our understanding
of spin coherence in small mesoscopic systems and more importantly, 
at or across interfaces (metal/semiconductor, 
semiconductor/semiconductor) is currently rudimentary to non-existent.
Much work (both theoretical and experimental as well as
materials and fabrication related) is needed to develop 
a comprehensive understanding of spin coherence in electronic
materials before the spintronics dream can become a viable reality.

\vspace{1cm}
\noindent

{\it Acknowledgments}--This work is supported by the U.S. ONR
and the DOD. We thank P. B. Allen and M. Johnson for useful 
discussions.

\begin{figure}
\caption{The relevant spin relaxation mechanisms for conduction
electrons. (A) {\it The Elliott-Yafet mechanism}. The periodic
spin-orbit interaction makes the spin ``up'' Bloch states contain
small spin down amplitude, and vice versa. Impurities, boundaries,
and phonons
can induce transitions between spin ``up'' and ``down'' leading
to spin degradation. (B) {\it The D'yakonov-Perel' mechanism}. In
noncentrosymmetric crystals spin bands are no longer degenerate:
in the same momentum state spin up has different energy than spin
down.  This is equivalent to having internal magnetic fields, one for
each momentum. The spin of an electron precesses along such a field,
until the electron momentum changes by impurity, boundary, or phonon
scattering. Then the precession starts again, but along a different
axis. Since the spin polarization changes during the precession,
the scattering acts against the spin relaxation. (C) {\it The
Bir-Aronov-Pikus mechanism}. The exchange interaction between
electrons and holes causes the electron spins  to precess along
some effective magnetic field determined by hole spins.
In the limit of strong hole spin relaxation, this effective
field randomly changes before the full precession is completed,
reducing the electron spin relaxation.
}
\label{fig:1}
\end{figure}

\begin{figure}
\caption{Revised Monod-Beuneu scaling. The measured width
$\Gamma_s=const\times (1/T_1)$ of the conduction electron spin
resonance signal is divided by the effective spin-mixing probability
$b^2$ obtained from atomic parameters, and by resistivity $\rho_D$ at
Debye temperature $T_D$. This should follow a Gr\"{u}neisen curve
when plotted as function of reduced temperature $T/T_D$. The alkali
metals fall onto a single curve while Al, Pd, Be, and Mg do not,
indicating that their $b^2$ is much larger than estimated
from atomic parameters.
}
\label{fig:2}
\end{figure}

\begin{figure}
\caption{The spin-hot-spot model. (A) {\it Monovalent metals}.
As electrons scatter and change momentum, they perform a random walk
on the Fermi surface. At each jump the electrons have a small
chance of flipping their spin (the Elliott-Yafet mechanism),
indicated on the right. In monovalent metals this chance is
uniform over the Fermi surface and is roughly equal to
$(\lambda/E_F)^2$.
(B) {\it Polyvalent metals}. Fermi surfaces of polyvalent metals
contain spin hot spots (black stains),  which are states  at
Brillouin zone boundaries, special symmetry points, or
accidental degeneracy lines. If an electron jumps into such a state,
its chance of flipping spin is much enhanced.
 Although spin hot spots form a small part of the Fermi
surface and the probability that an electron jumps there is quite
small, they nevertheless dominate the electron spin relaxation.
}
\label{fig:3}
\end{figure}

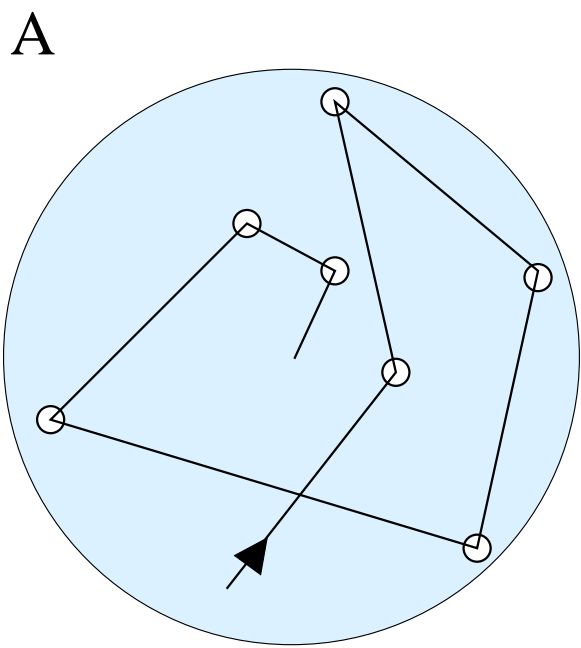
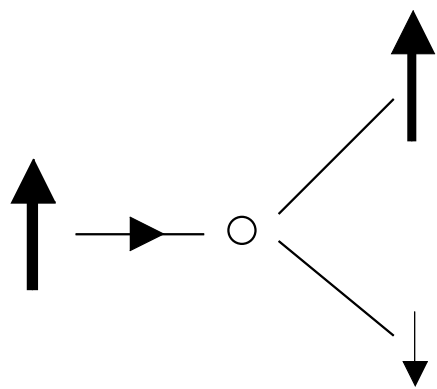
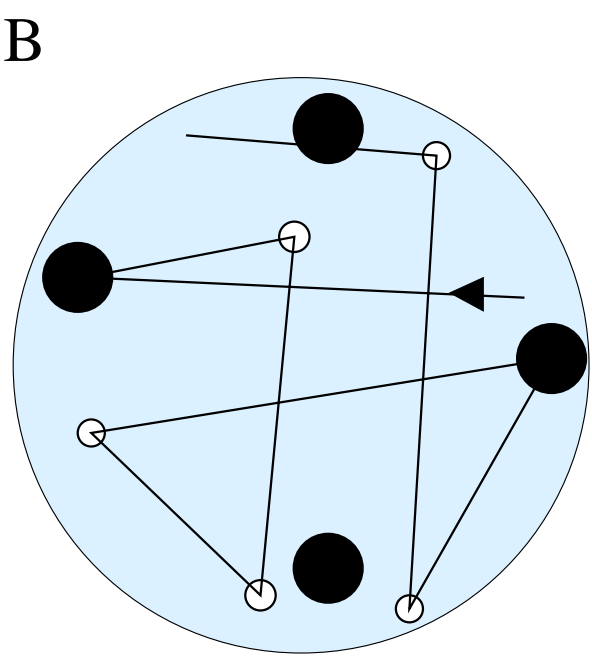
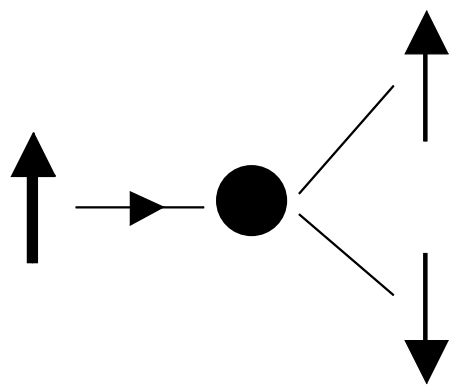
\end{document}